**Title:** Interpreting multi-variate models with setPCA

**Authors:** Nordine Aouni, Luc Linders, David Robinson, Len Vandelaer, Jessica Wiezorek, Geetesh Gupta, Rachel Cavill*

**Affiliation:** Department of Data Science and Knowledge Engineering, Maastricht University, The Netherlands.

*rachel.cavill@maastrichtuniversity.nl

# Interpreting multi-variate models with setPCA

**Nordine Aouni, Luc Linders, David Robinson, Len Vandelaer, Jessica Wiezorek, Geetesh Gupta, Rachel Cavill***

**Department of Data Science and Knowledge Engineering, Maastricht University, The Netherlands.**

*rachel.cavill@maastrichtuniversity.nl

Abstract

Principal Component Analysis (PCA) and other multi-variate models are often used in the analysis of "omics" data. These models contain much information which is currently neither easily accessible nor interpretable. Here we present an algorithmic method which has been developed to integrate this information with existing databases of background knowledge, stored in the form of known sets (for instance genesets or pathways). To make this accessible we have produced a Graphical User Interface (GUI) in Matlab which allows the overlay of known set information onto the loadings plot and thus improves the interpretability of the multi-variate model. For each known set the optimal convex hull, covering a subset of elements from the known set, is found through a search algorithm and displayed. In this paper we discuss two main topics; the details of the search algorithm for the optimal convex hull for this problem and the GUI interface which is freely available for download for academic use.

# 1 Introduction

Multi-variate models, in particular Principal Component Analysis (PCA) (Pearson, 1901) are a common part of the analysis pipeline for omics data. They allow the visualisation of the variance in the data projected into two or three dimensions, through the principal components (PCs). They can be used for clustering samples (Yeung & Ruzzo, 2001) or detecting problems (outliers, batch effects, confounding variables etc.) during the pre-processing pipeline (Eijssen et al., 2013). Once a multi-variate model has been built there are many different plots which can be made. The most common of these is the scores plot, which shows the relative positions of the samples in the model. There is an analogous plot for the variables, called the loadings plot, which shows the weights given to each variable in each PC. When fewer variables are used in building the model, for instance in NMR metabolomics, it is common to show the loadings plot alongside the scores plot to aid the biological interpretation of the separation/clustering seen in the scores plot e.g. (Clayton et al., 2006). However, when thousands (or more) variables are used in the model these plots become too dense and therefore uninterpretable. Additionally, the loadings will be formed from the interaction of thousands of variables acting in concert, so even if the values are examined, the information contained in these values is too dense to be clearly interpreted manually.

Hence, much information contained in these models is not currently exploited. When considering transcriptomics data each point in our loadings plot represents a gene. Through existing databases we have information on the pathway, functional and other groupings for many of these genes and by overlaying this information onto the loadings plot, we can obtain additional interpretable information, which will add value to the multi-variate models we build.

There has been some work in this direction; at the simplest level several publications have manually highlighted sets of genes or metabolites with a colour scheme in their loadings plots, for instance (Bylesjö et al., 2009; Lehrmann et al., 2012). More notably, Florian Wagner produced GO-PCA, a method which looks at the enrichment of Gene Ontology (GO) terms when considering the weights of the genes in the PCA model (Wagner, 2015). Similar ideas have been used to interpret PCA models of metabolomics data (Yamamoto et al., 2014). Another related method is PLIER (Mao et al., 2019) which is a matrix factorization method that generates a latent variable model of the data based on a sparse selection of pathways whose genes have weights in each latent variable. Principal component gene set enrichment (PCGSE) (Frost et al., 2015) is another approach to this problem implemented in an R Package which allows for three different measures of association between the gene and the PC to be tested; the raw loadings; the Pearson correlation co-efficient between the expression and the loading of interest and a Fisher transformed correlation co-efficient. It implements two enrichment tests, a mean difference statistic and a Wilcoxon signed rank test, with three measures of significance, parametric, correlation-adjusted parametric and a standard permutation.

However, our method goes further than simply identifying and listing the enriched sets. We have developed algorithms to overlay them onto the PCA loadings plot and display polygons which enclose the optimal group of elements from each set. Through the development of a Matlab GUI environment we present this information in an interactive format, with adjustable thresholds for the sets which are displayed. An alternative implementation in python has also been developed.

## 2 Loadings plots in multi-variate models

When omics practitioners apply a multi-variate model, such as a PCA, PLS or PLS-DA, the usual plot shown is the scores plot. In the scores plot each sample is a single point. The position of the point in the plot is determined by a linear weighted sum of the variables. The weights in this sum are the same for every sample. Therefore, we can produce an analogous plot which displays one point per variable, where the position of the variable is determined by its weight in the weighted sum for each latent variable (for instance each principal component in PCA). We call this plot the *loadings plot*. This plot should be interpreted in conjunction with the scores plot. A sample which is in the top right hand corner of the scores plot, will have a high value in many of the variables which are in the top right of the loadings plot.

If the multi-variate model has been built using a small number of variables then it is simple to label the points in the loadings plot and interpret the output. This is usual in many NMR-based metabolomics studies (Ringeissen et al., 2003; Schnackenberg et al., 2006; Selman et al., 2006). However, once the number of variables becomes large then this becomes impractical.

However, these models and plots still contain much unutilised information. Variables which are grouped together will show similar patterns of expression across the samples and therefore, if these groups match known sets of variables then this is of great interest as it will lead to a greater interpretability of the models produced.

Therefore, we present an algorithmic approach to integrate databases which contain information on sets of variables, for instance, sets of genes with these multi-variate models and present the results.

# 3   Convex hull optimization algorithm

The algorithm takes as input a list of known sets and the weights of each variable in two selected latent variables from the multi-variate model. In the first stage of the algorithm the optimal convex hull is calculated for each gene set in the list. In the next stage, various filter criteria are calculated, which can then be used to limit which convex hulls are displayed. The third stage is visualisation.

## 3.1   Optimal Convex Hull calculation

The convex hull around a set a points is a unique polygon which covers all points given and to travel in a straight line between any two points under the convex hull one would not have to leave the convex hull. Formally, this can be defined in terms of a weighted sum of points, where all weights are all non-negative and sum to one, as shown in equation 1 below.

$$Conv(S) = \left\{ \sum_{i=1}^{|S|} a_i x_i \mid (\forall i : a_i \geq 0) \wedge \sum_{i=1}^{|S|} a_i = 1 \right\} \qquad (1)$$

Where $a_i$ is the weight for a particular given point $x_i$.

Convex hulls were chosen as the shapes to overlay a set of variables in the loadings plot as they have a useful feature, namely that the removal of any point from the set generating the convex hull, except corner points, will result in the same convex hull being obtained. This means that the branching factor in the search for the optimal convex hull is the same as the number of corner points on the polygon and therefore not all combinations of points from the set need to be considered.

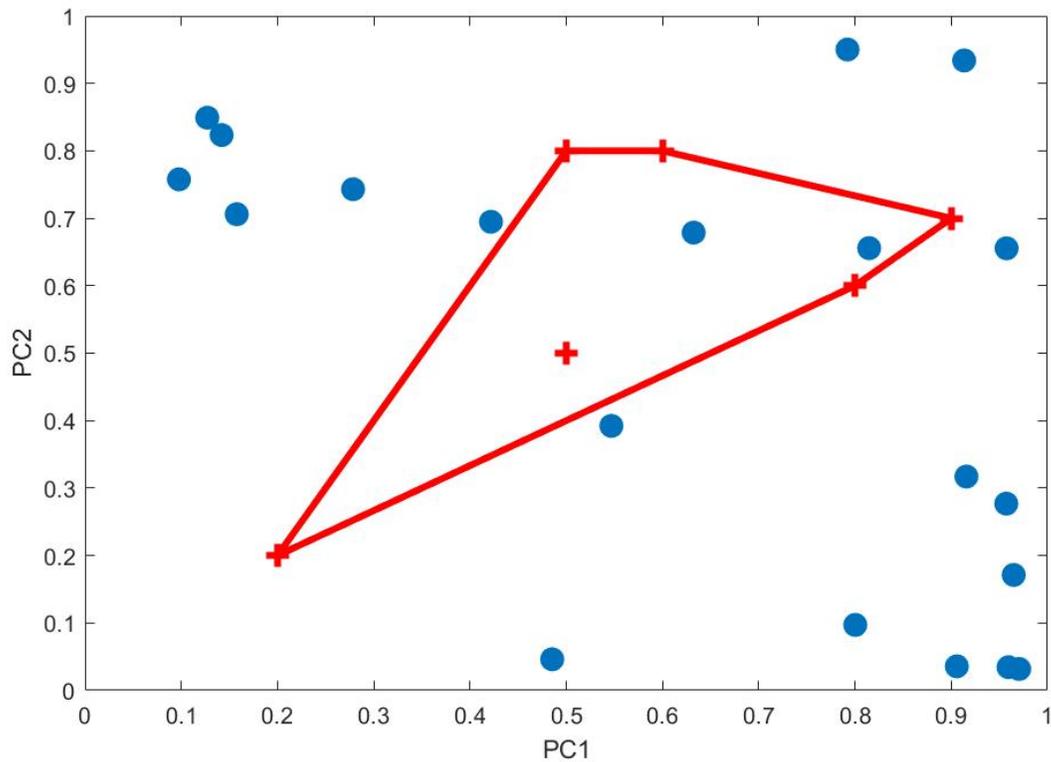

*Figure 1 A loadings plot with a convex hull enclosing elements of a known set shown in red superimposed. Other variables are shown with blue circles. Note: the variable marked by the red cross in the middle of the shape can be removed from the set without changing the convex hull.*

Having defined the convex hull, we next need to define what we mean by the optimal convex hull. For this, consider the situation shown in figure 1. Many of the variables from the set are grouped together in one part of the loadings plot, indicating that they have similar expression across the samples, however there is one element of the known set which lies away from the rest. We wish therefore, to exclude this element from the convex hull[1]. To do this we optimise the over-representation of points from the set under the convex hull. More formerly, given a known set of elements P, we can generate a convex hull $H_P$ which covers those points. We also have the set of all variables G and we define |P| = n, |G|=N, S={g|g in $H_P$} and |$H_P$|=k. In words, this means that our known set has size n, our entire variable list has size N, we have k elements lying underneath the convex hull and we have another set S which contains the overlap of P and G which are under the convex hull, so all the elements in the known set which are covered by the current convex hull. Then using the standard Fisher's over-representation formula (Fisher, 1925) we can calculate the score of this convex hull.

---

[1] This process of removal is similar in intention to gene-set reduction (Dinu et al., 2009), although achieved through a very different method.

$$f(H_P) = \sum_{x=|S|}^{\min(n,k)} \frac{\binom{k}{x}\binom{N-k}{n-x}}{\binom{N}{n}} \quad (2)$$

## 3.2  Searching for the optimal convex hull

Having defined a score for each convex hull as above, we can use a search algorithm to find the optimal convex hull for a given known set. We can perform this search either in a top-down or a bottom-up manner.

In the top down search we start with all the variables from the known set being under the convex hull and then expand the search tree by removing corners (see figure 2). Using this approach we will sometimes find sets which have already been considered in other branches of the tree, we can prevent these being re-examined by keeping a record of which combinations of elements we have already considered and pruning these branches.

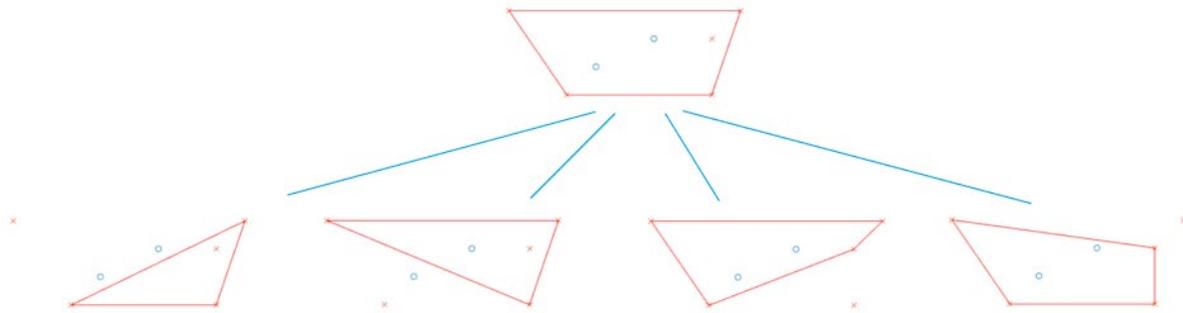

*Figure 2 Top down tree search of convex hulls. Points in the set are shown with red crosses, points outside the set are shown with blue circles. To search through the possible hulls one excludes one corner point from the original hull to generate a new search branch.*

Trees can also be built using a bottom-up approach. Here the root node will contain no points and children will be generated by adding one element that is not contained inside the convex hull of the current search node. However, this can result in other elements from the known set being included inside the convex hull of the new node that were not present in the parent node. This means search nodes at equal depth will have different numbers of gene points from the pathway in them. When calculating node score, the algorithm looks at which points are within the hull. Any new element will be added from those not inside the hull of the current search node to ensure the convex hull will change. The tree search algorithm will then explore all the nodes searched with the top down approach, plus the additional nodes with less than three gene points at the top of the tree. The number of additional nodes to be searched is insignificant for larger trees and as convex hulls do not need to be computed for them, the run-time using this approach does not increase by any significant amount.

### 3.2.1 Theoretical limits of the tree search

In this section we are going to explore the theoretical best and worst case scenarios that can be found during our tree search for the optimal convex hull. By understanding this we can effectively prune the tree and hence speed up the search. However, this section is not necessary to understand the use of the algorithm and can be skipped by readers who are not interested in the mathematical properties of our search.

Using the top-down approach, the worst score for a child node $c$ of parent node $p$ would be when all non-pathway points in the convex hull of $p$ are also included inside the convex hull of $c$ (see figure 3a). The best score would be when no non-pathway points are inside the convex hull of $c$, see figure 3b. This gives the bounds for the score of the child node as:

$$\frac{\binom{N-|S|_p+1}{n-|S|_p+1}}{\binom{N}{n}} \leq c \leq \sum_{x=|S|_p-1}^{\min(n,k_p-1)} \frac{\binom{k_p-1}{x}\binom{N-k_p+1}{n-x}}{\binom{N}{n}} \qquad (3)$$

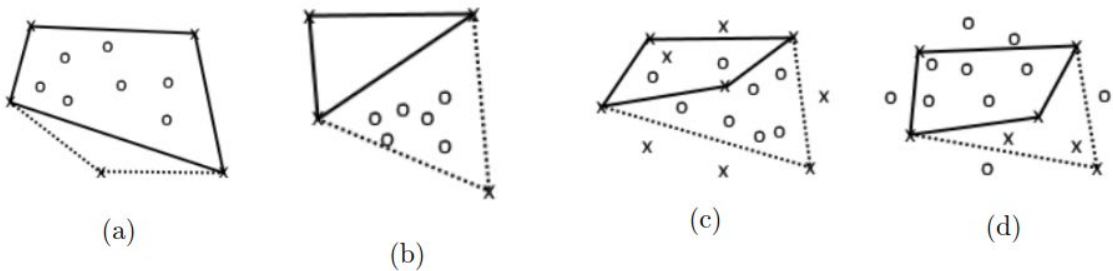

(a) (b) (c) (d)

Figure 3 Convex hulls, crossed points are in the known set P, empty circles are in G but not in P. The dotted line shows the convex hull of the parent node in the search and the solid line shows the convex hull of the child node. a) Top down worst case b) Top down best case c) Bottom up worst case and d) Bottom up best case.

Using the bottom-up approach, the worst score for a child node *c* of parent node *p* would be when all non-pathway points in the dataset are included inside the convex hull of *c* with only one true point added (see figure 3c). The best score would be when no additional non-pathway points are inside the convex hull of *c* but all pathway points are added, see figure 3d. The best case also holds for all descendants of *p*. This gives the bounds:

$$\frac{\binom{k_p+n-|S|_p}{n}}{\binom{N}{n}} \leq c \leq \sum_{x=|S|_p+1}^{\min(n,N-n+|S|_p+1)} \frac{\binom{N-n+|S|_p+1}{x}\binom{n-|S|_p-1}{n-x}}{\binom{N}{n}} \quad (4)$$

### 3.2.2 Speeding up the search

Running the complete search described above is a computationally intensive task, especially for large known sets, for instance GO-terms which can have thousands of member genes. Therefore, we employ some techniques to speed up the search.

A greedy best first search will evaluate all children of a search node and then select the best one (in our case the lowest score) to expand and search first. Added to this we can use the limits discussed in the previous section to apply pruning. When performing bottom-up search as the left hand part of equation 4 holds not just for the direct child of the node, but also for any descendant of the search node, we can safely prune any child where their score is larger than the limit. A node and all its successors can also be pruned if we have already found a score lower than the upper bound for that node.

## 4 Software

Having described the approach and the convex hull search algorithm above we will now describe the software in which these are implemented. This software is called GenesetPCA and is implemented as a Matlab GUI available alongside a full instruction manual from

[https://dke.maastrichtuniversity.nl/rachel.cavill/](https://dke.maastrichtuniversity.nl/rachel.cavill/).

From the same site there is also a python implementation available, although the GUI is less fully featured than the Matlab GUI described here.

### 4.1 Data input

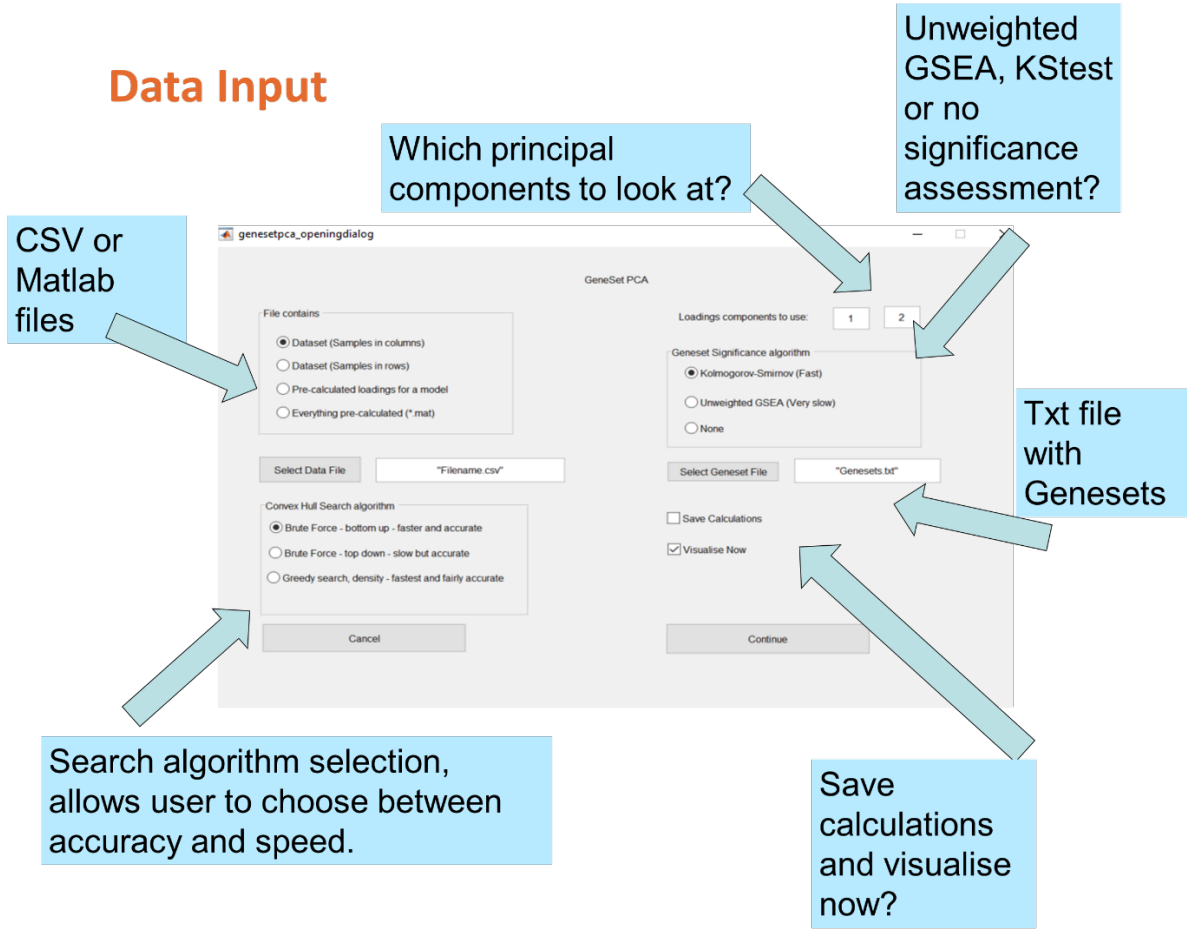

*Figure 4 The data input screen explained.*

The process of running the software starts with data input. The data input screen is shown in figure 4, here the user can select between the different algorithms offered.

The first box is the 'File contains' box, here the user should select between uploading a dataset or pre-calculated loadings from a model. Although the method can be applied to any type of multi-variate model at this point in time only PCA models are calculated by the program internally. To use the method with other types of linear multi-variate models (for instance PLS), one has to first calculate the model based on your data and then upload the loadings into our program.

The second option box allows the user to specify the algorithm they wish to use for the search.  As detailed above, different levels of pruning can be applied to the search to speed it up, however some of these options may reduce the accuracy in identifying the optimal convex hull for every known set.

In the second column the user can select which principal components they are interested in examining.  For instance, if the separation between the groups of interest was found in PC3 and PC5 then these should be entered here.

Next we can select the enrichment algorithm applied, choosing between unweighted GSEA or a Kolmogorov-Smirnov test.  These algorithms will be described in more detail in the following section.

Finally, there are options to select the file containing the known sets, to save the calculations made and to automatically start the visualisation in the GUI.

## 4.2   Calculating measures for display filtering

At this point the software will now calculate the optimal convex hull using the options selected above.  Once the optimal convex hull for each known set has been calculated we can calculate various properties of this hull. These values will allow us to effectively filter which hulls are displayed in an interactive GUI.

Some of these values are p-values and as we are testing many known sets we correct these p-values using Benjamini-Hochberg False Discovery rate (FDR) to correct for multiple testing (Benjamini & Hochberg, 1995).

*Table 1 Filter values calculated per convex hull.*

| Value | Type | Range | Notes |
|---|---|---|---|
| Convex Hull significance | FDR corrected p-value | $0 < x \leq 1$ | Pre-calculated in the search |
| Set size | Ordinal | $3 \leq x < 1000+$ | Size of the known set |
| Minimum percentage of genes from known set in the convex hull | Percentage | $0 < x \leq 100$ | |
| 1st component significance | FDR corrected p-value | $0 < x \leq 1$ | Enrichment calculation |
| 2nd component significance | FDR corrected p-value | $0 < x \leq 1$ | Enrichment calculation |

| Loadings angle significance | FDR corrected p-value | 0 < x ≤ 1 | Enrichment calculation |

In table 1 the first three values are either already calculated during the search process or can be trivially calculated from the values therein. The last three filter values calculated are generated from an enrichment calculation.

### 4.2.1 Enrichment calculations

The concept behind an enrichment calculation is to take an ordered list of elements and to calculate the likelihood of elements from a known set grouping in the way that is seen in the data. We can use different algorithms to calculate enrichment. Two common algorithms for this task are unweighted Gene Set Enrichment Analysis (GSEA) (Subramanian et al., 2005) and the Kolmogorov-Smirnov test (Kolmogorov, 1933; Smirnov, 1948).

In this application we selected the unweighted GSEA as opposed to the weighted version where elements are weighted to have more influence on the score if they have higher values (Subramanian et al., 2007). We do this as we want to identify clusters of elements from the known set regardless of where they sit on the axis, a cluster in the middle of the plot may be as interesting as one at the edges. The disadvantage of GSEA is that it generates its p-values through permutations and therefore the computational cost of generating accurate p-values is very high as it is known that to get accurate results many permutations are necessary (Keller et al., 2007).

An alternative to the unweighted GSEA is the Kolmogorov-Smirnov test. Here one looks at the cumulative distributions of the elements in the known set, compared with the same distribution generated from all elements. The test then evaluates the significance of the point at which these distributions diverge the most. This is a very fast test to perform.

We apply these enrichment algorithms in three cases, firstly in the direction of the x-axis on the loadings plot, secondly in the direction of the y-axis on the loadings plot and thirdly with the loadings angle.

The loadings angle enables us to combine both the x and y directions. We take the angle of the line between each element and the origin and look for enrichment in these values. However, as this is a circular list and our enrichment algorithms assume a linear list we must cut it somewhere. In order to not miss known sets that are inadvertently cut in two by this process we perform two enrichments with cuts opposite each other in the circle. For each known set the minimum value that it scores in either of these enrichments is taken as its significance.

### 4.2.2 Removing redundant known sets

Many databases of known sets will contain extensive overlap between pairs of sets. For instance, if we consider GO-terms as our known sets, then these have a nested structure where a parent term will contain at least all elements of all their children. Given our approach to finding the optimal convex hull, when we consider nested or overlapping sets we may end up finding the same subset of elements for multiple known sets as their optimal convex hulls. This does not help the user. Therefore, we have found that implementing some simple rules allows us to eliminate this redundancy.

Firstly, we order all known sets according to the significance of their convex hulls. In the case where multiple known sets result in the same convex hull, this value will still differ, depending on the initial size of the known set and the number of elements covered by the convex hull. In general, this will favour small sets or those with more elements covered by the convex hull.

We start by taking the most significant known set, then we descend this ordered list until we find a known set which has zero overlap with this set, this then gets added to our results list. We proceed incrementally, only adding sets to the results list which have no overlap with any of the sets currently in the results list. This is a very stringent strategy, but it has proved efficient in narrowing down the results list to a much more manageable number of results. A second, even stricter strategy is to look for convex hulls which don't overlap in area.

### 4.3 Interactive GUI

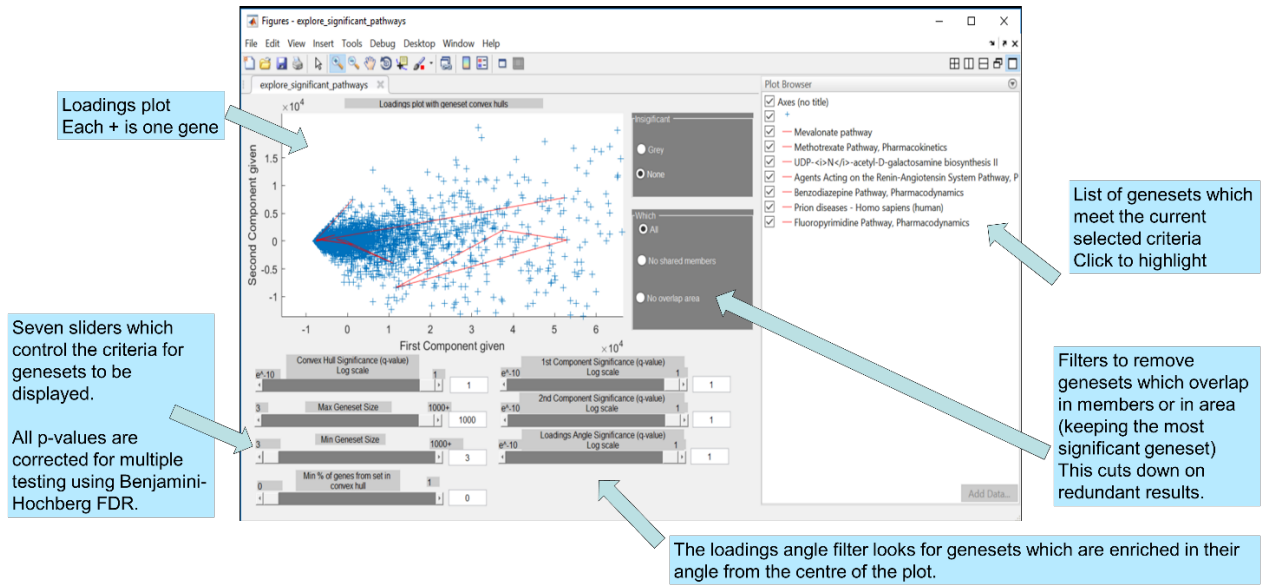

*Figure 5 Typical output in the visualisation GUI for transcriptomics data.*

Having calculated not only the optimal convex hull for each pathway, but also the filtering values it is now time to run the GUI as shown in figure 5. Here we have an interactive environment in which users can explore the known sets which can add interpretative value to their model. In order to facilitate this process we have included sliders to filter on all the filter values and radio buttons to activate the removal of redundant results as described above. By adjusting the sliders and using the radio buttons users will find combinations which give them the most interesting pathways for their datasets without relying on generic thresholds or heuristics as to the best values for these filters. For instance, in one dataset the enrichment in PC1 may be very important, yet in another PC1 may capture variation not related to the feature of interest and therefore this filter could be set to 1.

Implementing this as a Matlab GUI meant that many of the native matlab features, such as the plot browser window could be exploited to give the user the additional information they require – such as the name of each known set displayed. Zooming, moving and data point labelling are also natively integrated into the GUI.

# 5 Conclusion

In this paper we have presented a both an algorithmic approach for identifying known sets of elements which strongly influence a multi-variate model and a software implementation of said algorithm. We believe that exploring the data in this way can enable omics practitioners to interpret their multi-variate models and hence increase the information retrieval and hypothesis generation possible from these models.